\documentclass[10pt,journal]{IEEEtran}
\ifCLASSOPTIONcompsoc
\usepackage[nocompress]{cite}
\else
\usepackage{cite}
\fi

\usepackage{booktabs}   
\usepackage{amsthm}
\usepackage{amsmath}
\usepackage{mathtools}
\usepackage{setspace}
\usepackage{algorithmic}
\usepackage{graphicx}
\usepackage{textcomp}
\usepackage{verbatim}
\usepackage[table]{xcolor}
\usepackage{soul}
\usepackage{makecell}
\usepackage{amsmath}
\usepackage{cite}

\makeatletter

\renewenvironment{proof}[1][\proofname]{%
\par\pushQED{\qed}\normalfont%
\topsep6\p@\@plus6\p@\relax
\trivlist\item[\hskip\labelsep\bfseries#1\@addpunct{.}]%
\ignorespaces
}{%
\popQED\endtrivlist\@endpefalse
}

\makeatother

\newtheoremstyle{mytheoremstyle} 
{}                    
{}                    
{}                   
{}                           
{\bfseries}                   
{:}                          
{0.5em}                       
{}  

\theoremstyle{mytheoremstyle}
\newtheorem{identity}{Identity}

\newcommand{\iandSymbol}{\includegraphics[width=0.70em]{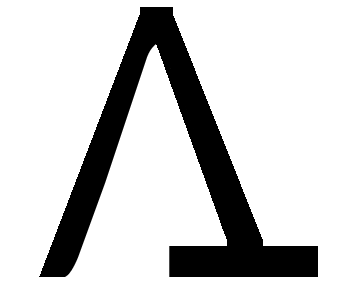}}

\newcommand{\iand}[3]{(#1\: \iandSymbol\: #2)\: \iandSymbol\: #3}
\newcommand{\iandtwo}[2]{#1\: \iandSymbol\: #2}

\newcommand{\piand}[3]{((#1\: \iandSymbol\: #2)\: \iandSymbol\: #3)}
\newcommand{\piandtwo}[2]{(#1\: \iandSymbol\: #2)}

\newcommand{\andthree}[3]{#1\wedge#2\wedge#3}
\newcommand{\andtwo}[2]{#1\wedge#2}

\newcommand{\pandthree}[3]{(#1\wedge#2\wedge#3)}
\newcommand{\pandtwo}[2]{(#1\wedge#2)}

\newcommand{\veryshortarrow}[1][3pt]{\mathrel{%
   \hbox{\rule[\dimexpr\fontdimen22\textfont2-.2pt\relax]{#1}{.4pt}}%
   \mkern-4mu\hbox{\usefont{U}{lasy}{m}{n}\symbol{41}}}}
   
\newcommand{\imp}[3]{#1 \veryshortarrow (#2 \veryshortarrow #3)}
\newcommand{\imptwo}[2]{#1\rightarrow#2}
\newcommand{\pimptwo}[2]{(#1 \rightarrow #2)}

\newcommand{\pimp}[3]{(#1 \veryshortarrow (#2 \veryshortarrow #3))}

\newtheorem{theorem}{Theorem}

\newcommand{\iandoption}{\mathrm{(a)}\:\mathrm{IAND:\;\;}}
\newcommand{\implyoption}{\mathrm{(b)}\:\mathrm{IMPLY:\;\;}}

\begin{document}

\title{Complete Boolean Algebra for Memristive and Spintronic Asymmetric Basis Logic Functions}
\graphicspath{{figures/}}
\author{Vaibhav Vyas and Joseph S. Friedman, \IEEEmembership{Senior Member, IEEE}
\thanks{The authors are with the Department of Electrical and Computer Engineering, University of Texas at Dallas, Richardson, TX 75080 USA.}
\thanks{Corresponding author: Vaibhav Vyas.}
\thanks{Email: vyasvaibhav2@gmail.com, joseph.friedman@utdallas.edu.}}

\IEEEtitleabstractindextext{%
\begin{abstract}
The increasing advancement of emerging device technologies that provide alternative basis logic sets necessitates the exploration of innovative logic design automation methodologies. Specifically, emerging computing architectures based on the memristor and the bilayer avalanche spin-diode offer non-commutative or `asymmetric' operations, namely the inverted-input AND (IAND) and implication as basis logic gates. Existing logic design techniques inadequately leverage the unique characteristics of asymmetric logic functions resulting in insufficiently optimized logic circuits. This paper presents a complete Boolean algebraic framework specifically tailored to asymmetric logic functions, introducing fundamental identities, theorems and canonical normal forms that lay the groundwork for efficient synthesis and minimization of such logic circuits without relying on conventional Boolean algebra. Further, this paper establishes a logical relationship between implication and IAND operations. A previously proposed modified Karnaugh map method based on a subset of the presented algebraic principles demonstrated a 28\% reduction in computational steps for an algorithmically designed memristive full adder; the presently-proposed algebraic framework lays the foundation for much greater future improvements.  
\end{abstract}

}
\maketitle

\IEEEdisplaynontitleabstractindextext

\IEEEpeerreviewmaketitle


\section{Introduction}
\label{sec:intro}
Over the last five decades, progress in fabrication technology has enabled the continued downscaling of CMOS-based devices that has contributed to consistent improvements in the performance of integrated circuits (ICs) in accordance with Moore's Law \cite{kim2003leakage, lundstrom2022moore}. However, achieving further scaling becomes exceedingly challenging as the transistor device size approaches a few atomic layers, triggering the onset of quantum effects \cite{bohr2017cmos, waldrop2016chips,finocchio2023roadmap}. This may result in heightened leakage current caused by quantum tunneling, consequently raising the standby power dissipation \cite{liu2019hybrid, lin2019two}. Fortunately, in the pursuit of finding alternative technologies, several promising proposals have been put forth, including magnetic tunnel junction logic \cite{bromberg2012novel, cmatTED}, domain wall logic \cite{allwood2005magnetic, omari2014chirality, dml3,dml4_mqca,bromberg2012novel}, all-carbon spin logic \cite{allcarbon}, reversible skyrmion logic \cite{walker2023near}, spin-FETs \cite{datta1990electronic, spinfet2_pre,spinfet3_nb,spinfet4_q} that use magnets and electron spin as state variables instead of charge currents, and other spin based logic systems \cite{barla2021spintronic, basdljf, vyas2018sequential}. 

As newer devices emerge, it is imperative that progress in logic design matches the pace of device research through refinement of existing logic synthesis and minimization techniques that suit these novel device technologies. Specific device challenges and opportunities include the bilayer avalanche spin-diode device \cite{basdljf} can be used to perform logical OR or an inverted-input AND (IAND) function, while the memristor-based stateful logic \cite{kvatinsky2011memristor, Kvatinsky2014} efficiently performs the implication (IMPLY) and NAND boolean functions. Each of the IMPLY and IAND boolean logic operations are logically `asymmetric', meaning that they are non-commutative in nature. In many previous studies, logic minimization is conducted using a basis set of traditional logic functions, whereby each foundational logic gate is mapped to an efficient memristor implementation. For instance, \cite{Chattopadhyay2011} minimizes a function into OR and inverter gates, while \cite{Raghuvanshi2014} utilizes NOT, NAND and OR gates for minimization. Most notably, \cite{Lehtonen2012} minimizes a substantial function into NAND, OR, and parity gates, subsequently implementing them with memristors. Further, \cite{Shirinzadeh2016} employs majority gates to achieve optimized circuits, and while \cite{Marranghello2015a} focuses on minimizing the number of memristors, it does not present a way to reduce the number of computational steps. Binary decision diagrams \cite{Chakraborti2014} and CMOS buffer circuits \cite{Lalchhandama2016} have also been applied, as well as an interpretation of memristors as threshold logic elements \cite{Fan2014}. Since all of these methodologies involve optimizing logic circuits using conventional symmetric Boolean algebraic principles, the resulting circuits fail to fully capitalize on the asymmetric basis logic operations offered by memristors. 

Previously, we have introduced a modified Karnaugh map method to minimize asymmetric logic circuits without the need to deploy conventional Boolean algebra \cite{vyaskmap}. A full adder circuit designed using the proposed minimization algorithm achieved a 28 percent reduction in the number of computational steps when compared to the best \cite{Kvatinsky2014} manually-optimized full adder. While the foundational Boolean algebraic principles underlying the modified Karnaugh map method were briefly outlined, this article presents a complete set of Boolean algebraic identities and theorems tailored to asymmetric logic functions. For completeness, this includes the principles that have already been laid out in our previous work. Section \ref{sec:background} details the background for the device technologies, asymmetric logic functions and the previously proposed Karnaugh map method for asymmetric logic, while Section \ref{sec:equations} presents the algebraic theorems and identities with necessary proofs. Section \ref{sec:conclusion} provides concluding remarks. 

\section{Background} 
\label{sec:background}
Both the bilayer avalanche spin-diode and memristor offer a distinct set of basis logic operations. A single bilayer avalanche spin-diode unit can perform both IAND and OR functions, while it takes two and three non-volatile memristors to perform an implication and NAND function, respectively. To efficiently leverage these basis logic functions in large-scale systems, it is crucial to acknowledge the logically asymmetric characteristics of the IAND and implication functions. Utilizing conventional logic design techniques to minimize such asymmetric logic functions yields a sub-optimal circuit, as the minimization process is not directly applied to the IMPLY-NAND and IAND-OR logic sets. Therefore, the development of a bespoke algebraic framework becomes imperative to fully harness the potential of such circuits.

\subsection{Bilayer Avalanche Spin-Diode Logic}
\label{sec:basdl_background}
\begin{figure}[t]
		\centering
		\includegraphics[width=0.70\columnwidth]{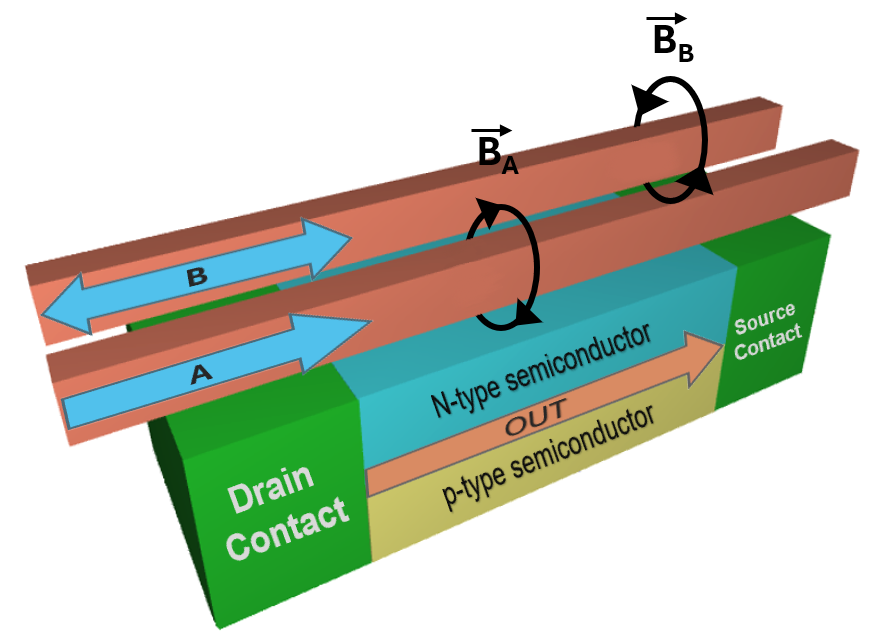}
		\caption{The bilayer avalanche spin-diode. The flow of currents in control wires A and B induces a magnetic field over the semiconductor p-n junction, leading to the modulation of the output current.}
		\label{fig:basdl}
\end{figure}

A bilayer avalanche spin-diode is a two-terminal spintronic device and possesses a negative magnetoresistance that allows an applied magnetic field to alter the resistance. The current passing through the spin-diode is modulated by the current on two control wires A and B (refer to Fig. \ref{fig:basdl}). The voltage across the diode is kept constant, enabling the two input currents to generate magnetic fields that affect the output current of the spin-diode. The resistance state of the spin-diode is dictated by the magnitude and alignment of the magnetic fields relative to a threshold level. 

\begin{figure}[t]
		\centering
		\includegraphics[width=0.8\columnwidth]{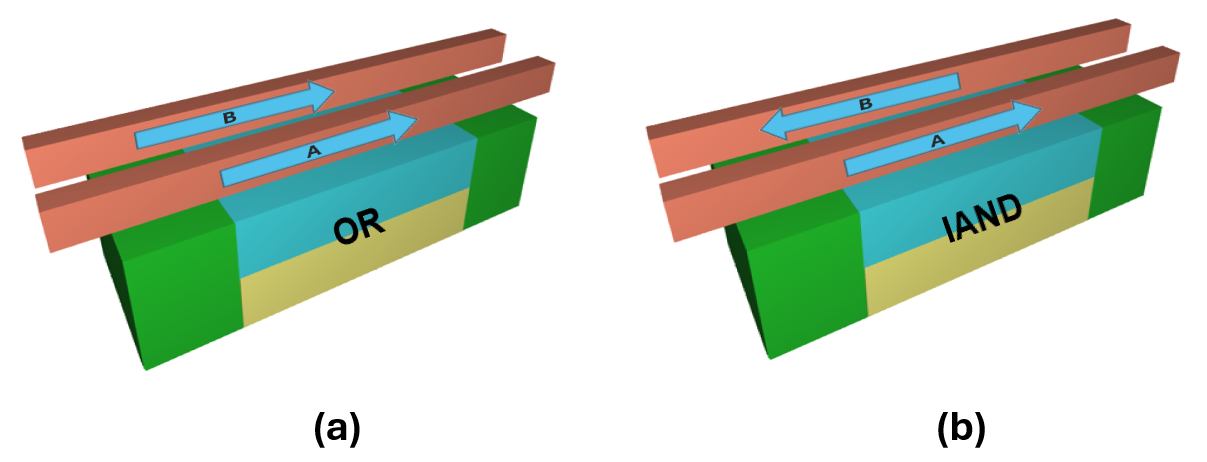}
		\caption{Bilayer avalanche spin-diode device performs a (a) Boolean OR if the input currents A and B flow in the same direction, and an (b) IAND function if A and B have opposite directions.}
		\label{fig:basdl_gates}
\end{figure}

In this system, a `1' is represented by a large current flow, while a `0' is represented by a small current. The device can perform a Boolean OR and an inverted-input AND (IAND) function depending on the relative direction of the control currents. These two functionalities arise from the magnetic fields generated by currents aligned in either the same or opposite directions, which can either reinforce or counteract each other, respectively  (see Fig. \ref{fig:basdl_gates} and Table \ref{tb:tt}). In contrast to memristors, spin-diodes exhibit volatility, reverting to their zero-magnetic field state promptly upon the cessation of applied input currents.


\subsection{Stateful Memristor Logic}
\label{sec:memristor_background}
\begin{figure}[t]
	\centering
	\includegraphics[width=0.75\columnwidth]{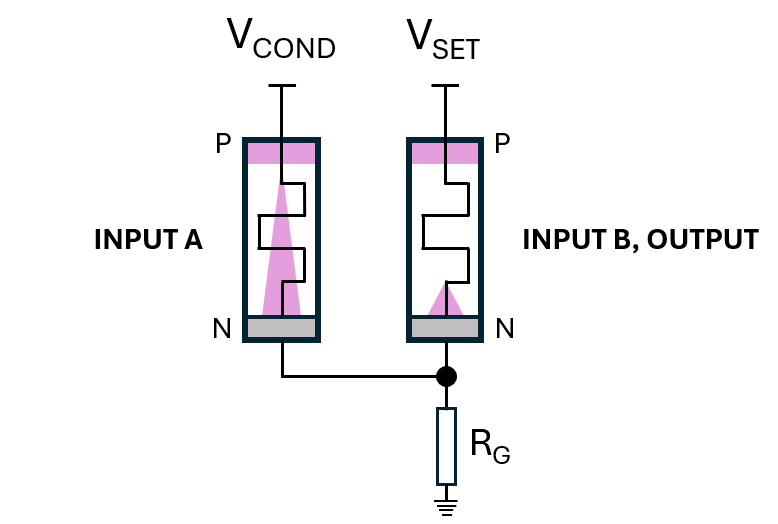}
	\caption{Schematic diagram illustrating the memristive implication logic, in which the resistance state is modulated by applying voltages to the memristors.}
	\label{fig:memristor}
	\vspace{-1.21em}
\end{figure}

Memristors, or `memory resistors' are two-terminal, non-linear electrical components whose resistance changes based on the amount of charge that has previously flown through them \cite{memristorchua, Strukov2009}.  First conceptualized by Leon Chua \cite{memristorchua} in 1971, the resistance of a memristor is not simply binary (\textit{i.e.}, in a high or low resistance state), but rather exists on a spectrum determined by the history of applied voltages. In an ideal scenario, when the applied voltage exceeds a certain threshold magnitude, it triggers a transition in the memristor's resistance state to a purely resistive or conductive state.
Specifically in Fig. \ref{fig:memristor}:
\begin{itemize}
    \item Conductive state (1): $V_{P}-V_{N} > V_{TH}$ and,
    \item Resistive state (0): $V_{N}-V_{P} > V_{TH}$,
\end{itemize}
where $V_{TH}$ is the threshold voltage, and $V_N$ and $V_P$ are the voltages at nodes $P$ and $N$ respectively.

 \begin{table}[t]
		\centering
		\bgroup
		\def\arraystretch{1}
		\caption{Truth table for implication and IAND Logic.}
 		\label{tb:tt}
		\resizebox{0.6\columnwidth}{!}{%
			\begin{tabular}{|c|c|c|c|}
				\hline
				\textbf{Input A} & \textbf{Input B} & \textbf{IMPLY} & \textbf{IAND} \\ \hline
				0                & 0                & 1             & 0              \\ \hline
				0                & 1                & 1             & 0              \\ \hline
				1                & 0                & 0             & 1              \\ \hline
				1                & 1                & 1             & 0              \\ \hline
			\end{tabular}%
		}
		\egroup
	\end{table}

\begin{figure}[t]
    \centering
    \includegraphics[width=0.65\columnwidth]{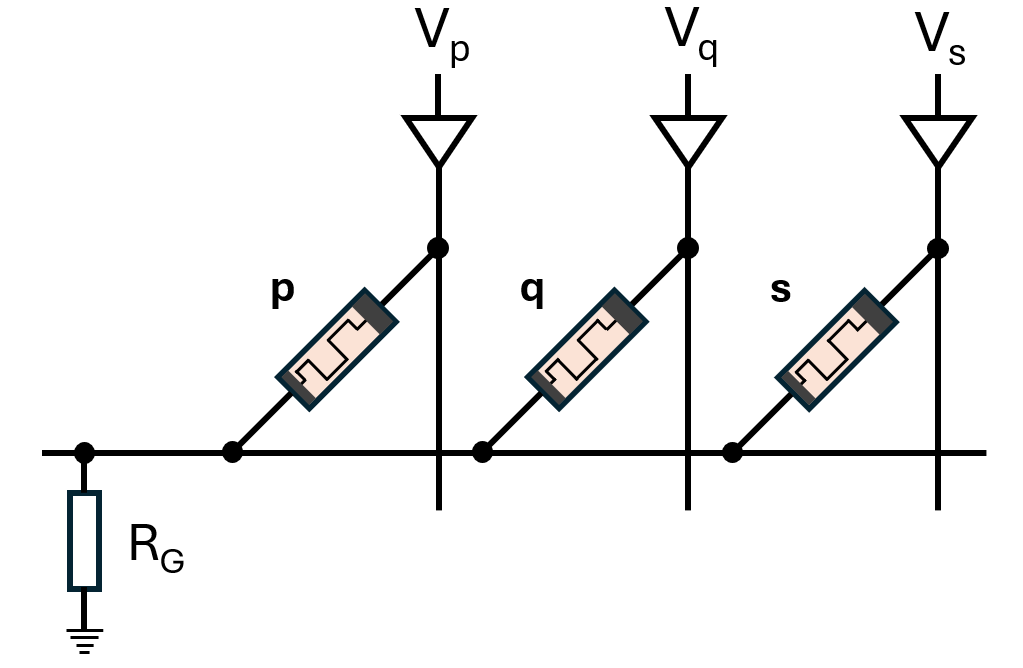}
    \caption{Standard NAND gate realization using stateful memristive implication.}
    \label{fig:imply_nand}
\end{figure}

Stateful implication, denoted as $OUTPUT = \imptwo{A}{B}$ is achieved by applying $V_{COND}$ to memristor $A$ and $V_{SET}$ to memristor $B$, maintaining $V_{COND} < V_{TH} < V_{SET}$ and $V_{SET} - V_{COND} < V_{TH}$ \cite{Borghetti2010}. Table \ref{tb:tt} shows the truth table for implication logic. When memristor $A$ is in the resistive state (0), the $V_{SET}$ voltage across memristor B exceeds $V_{TH}$, prompting memristor B to transition to the conductive state (1) or remain in it. Conversely, if memristor $A$ is in the conductive state (1), the voltage across memristor $B$ is $V_{SET} - V_{COND}$; since this value falls below $V_{TH}$, no switching occurs. Hence, the IMPLY function can be executed by two memristors in a single step, whereas a NAND function can similarly be executed with three memristors in two steps (see Fig. \ref{fig:imply_nand}). 

Unlike traditional approaches where each cascaded operation requires a distinct set of devices, in stateful memristor logic the memristors are continually reused through a sequential application of $V_{SET}$ and $V_{COND}$ voltages. Table \ref{tb:steps} outlines the sequential steps for implementing the NAND operation between $p$ and $q$, utilizing $s$ as the output memristor. 

\subsection{Asymmetric Basis Logic Functions}
\label{sec:asymm_func}
Typically, traditional Boolean operations adhere to commutativity, where the order of operands does not affect the outcome. However, asymmetric logic operations deviate from this norm as the result of such operations change when the operands are swapped. Thus, these operations can be characterized as inherently `non-commutative'. In this paper, the discussion centers around two such functions: IAND and IMPLY. 

The IAND function is an AND function with one input inverted (see Table \ref{tb:tt}). A distinct symbol (\iandSymbol) for this operation has been proposed in \cite{vyaskmap} such that
\begin{equation}
	IAND\:(A,B)=\andtwo{A}{\overline{B}}=\iandtwo{A}{B},
	\label{eq:bg1}
\end{equation}
whereas, implication operation is defined as
\begin{equation}
	IMPLY\:(A,B) = \overline{A} \vee B = \imptwo{A}{B}.
	\label{eq:bg2}
\end{equation}

Since in such operations the order of the operands is crucial, the operand to the left (right) side of the IAND (IMPLY) is called the \textit{non-inverted} input, while the remaining operand is called the \textit{inverted} input. Moreover, in scenarios involving more than two operands, only two operands must be computed at a time whilst paying attention to the overall order of operation. The following rules apply to any Boolean expression involving IAND or IMPLY operations:
	\begin{itemize}
		\item IAND: Two operands at a time, from left to right.
		\begin{equation}
		IAND\:(A,B,C)=\iand{A}{B}{C} 
		\label{eq:bg4}
		\end{equation}
		\item IMPLY: Two operands at a time, from right to left 
		\begin{equation}
		IMPLY\:(A,B,C)=\imp{A}{B}{C} 
		\label{eq:bg3}
		\end{equation}
	\end{itemize}

\subsection{Karnaugh Map Method for Asymmetric Logic Functions} 
\label{sec:bg_kmap}
The primary objective behind constructing an entire Boolean logic paradigm based on asymmetric logic sets is to streamline computation and optimization procedures for functions based on such sets, all the while eliminating the necessity to convert them into traditional logic. A comprehensively defined Boolean algebra tailored to asymmetric logic functions serves as the foundation for optimization algorithms, such as the modified Karnaugh map method proposed in \cite{vyaskmap}. The method is built upon a subset of theorems and identities presented in the subsequent sections of this paper, and enables the efficient design of a memristive full adder algorithmically designed using the proposed technique. The full adder was realized by using the IMPLY-NAND basis logic set, and the sum and carry out equations are denoted as NOI expressions as as shown in (\ref{eq:SumFA}) and (\ref{eq:CoutFA}) respectively. 

	\begin{figure}
		\centering
		\includegraphics[width=0.7\columnwidth]{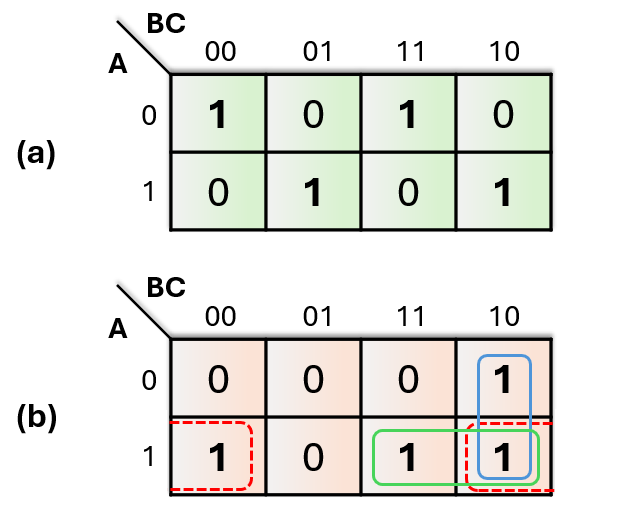}
		\caption{Karnaugh map representations for the full adder equations: (a) Sum (\ref{eq:SumFA}). (b) Carry out (\ref{eq:CoutFA}).}
		\label{fig:kmap_fa}
		\vspace{-1em}
	\end{figure}
 
\begin{multline}
	S_{noi}= \overline{\{(\imptwo{A}{(\imptwo{\overline{B}}{\overline{C}})})\wedge (\imptwo{\overline{A}}{(\imptwo{\overline{B}}{C})})} \\ \overline{\wedge(\imptwo{A}{(\imptwo{B}{C})})\wedge\ (\imptwo{\overline{A}}{(\imptwo{B}{\overline{C}})})\}}
	\label{eq:SumFA}
\end{multline}
\begin{multline}
	C_{noi}= \overline{\{(\imptwo{\overline{A}}{(\imptwo{B}{C})})\wedge (\imptwo{A}{(\imptwo{\overline{B}}{C})})} \\ \overline{\wedge(\imptwo{A}{(\imptwo{B}{\overline{C}})})\wedge\ (\imptwo{A}{(\imptwo{B}{C})})\}}
	\label{eq:CoutFA}
\end{multline}
The carry out expression is then reduced down to (\ref{eq:CoutFAred}) by using the suggested K-Map algorithm as shown in Fig. \ref{fig:kmap_fa}.  
	\begin{equation}
	C_{noi}= \overline{(\imptwo{A}{\overline{C}})\wedge (\imptwo{B}{\overline{C})\wedge (\imptwo{A}{\overline{B}})}}
	\label{eq:CoutFAred}
	\end{equation}
The resulting full adder realizes a 28\% reduction in the number of computational steps as compared to a previously proposed manual optimization approach \cite{Kvatinsky2014}. Moreover, this enhancement is even greater when applied to larger systems, such as multi-bit adders \cite{vyaskmap}.

\section{Boolean Algebra for Asymmetric Logic Functions}
\label{sec:equations}
The development of a Boolean algebra tailored specifically to asymmetric logic functions is necessary to enhance the efficiency of circuits employing devices that perform such functions inherently. This section provides the algebraic laws that can directly be applied to IAND and IMPLY functions without the need to represent them as conventional logic operations. Furthermore, canonical normal forms are presented to help standardize complex logical expressions involving these functions. Finally, a theoretical relationship between IAND and IMPLY functions is established. The laws governing IAND functions are accompanied by proofs where necessary, and while not explicitly proven for IMPLY functions, they can be demonstrated using a similar approach. 

\subsection{Core Algebraic Identities}
\label{sec:corealgebra}
The fundamental properties of a logic operation are defined by a set of core identities that are essential for simplifying Boolean expressions and ultimately for effective minimization of logic circuits. In this section, we present the essential algebraic identities for IAND and IMPLY functions.  

\subsubsection{Interaction with High (1) and Low (0) Logic}
These properties define how Boolean functions are altered when operated with IAND and IMPLY operators against a 0 or a 1. 

\begin{identity}[Annulment Law]
\label{th:annulmentlaw}
		\begin{equation}
		\iandoption\iandtwo{A}{1}\:=\:\iandtwo{0}{\overline{A}}\:=\:0 
	    \end{equation}
     	\begin{equation}
		\implyoption\imptwo{A}{1}\:=\:1 
	    \end{equation}
     
		\begin{proof}
			$\vspace{-0.5em}$
				Using AND representation for IAND, 
				\begin{equation}
				\iandtwo{A}{1}\:=\:\andtwo{A}{\overline{1}}\:=\:\andtwo{A}{0}\:=\:0
				\end{equation}
				\begin{equation}
				and, \;\;\;\iandtwo{0}{A}\:=\:\andtwo{0}{\overline{A}}\:=\:0
				\end{equation}
				proving the identity. 
		\end{proof}  
\end{identity}

\begin{table}[t]
		\bgroup
		\setlength{\tabcolsep}{0.3em}
		\def\arraystretch{1.5}
		\centering
		\caption{Computational steps for NAND logic realized via stateful memristive implication.}
		\label{tb:steps}
		\resizebox{\columnwidth}{!}{%
			\begin{tabular}{|c|c|c|c|c|c|c|}
				\hline
				\textbf{Step} & \textbf{Operation} &  \begin{tabular}[c]{@{}c@{}}$\mathbf{V_{COND}}$\\ \textbf{applied to}\end{tabular}& \begin{tabular}[c]{@{}c@{}}$\mathbf{V_{SET}}$\\ \textbf{applied to}\end{tabular}& \begin{tabular}[c]{@{}c@{}}$\mathbf{V_{RESET}}$\\ \textbf{applied to}\end{tabular} & \textbf{\begin{tabular}[c]{@{}c@{}}Output\\Memristor\end{tabular}} & \textbf{\begin{tabular}[c]{@{}c@{}}State\\of \textit{s}\end{tabular}} \\ \hline
				0    & RESET   & -      & -     & $s$      & -                                                           & $s=0$    \\ \hline
				1    & $\imptwo{p}{s}$          & $p$     & $s$    & -       &  $s$                                         & $\overline{p} $      \\ \hline
				2    & $\imptwo{q}{s} $         & $q$     &$s$    & -       &  $s$                                           & $\overline{p \wedge q}$      \\ \hline
			\end{tabular}
		}
		\egroup
	\end{table}
 
\begin{identity}[Inversion Law]
\label{th:inversionlaw}
        \begin{equation}
        \iandoption\iandtwo{1}{A}\:=\:\overline{A}
        \end{equation}
        \begin{equation}
        \implyoption\imptwo{A}{0}\:=\:\overline{A}
        \end{equation}	
        
	\begin{proof}
	Representing the IAND operation in terms of AND, 
		\begin{equation}
			\iandtwo{1}{A}\:=\:\andtwo{1}{\overline{A}}\:=\:\overline{A}
		\end{equation}
		which proves the identity.
\end{proof}
\end{identity}

\begin{identity}[Identity Law]
\label{id:identitylaw}
		\begin{equation}
		\iandoption\iandtwo{A}{0}\:=\:A	
		\end{equation}
            \begin{equation}
            \implyoption\imptwo{1}{A}\:=\:A
            \end{equation}
  
	\begin{proof}
		Representing the IAND operation in terms of AND, 
		\begin{equation}
		\iandtwo{A}{0}\:=\:\andtwo{A}{\overline{0}}\:=\:\andtwo{A}{1}\:=\:A
		\end{equation}
		which proves the identity.
	\end{proof}
\end{identity}

\subsubsection{Idempotency}
\label{sec:idempotency}
Idempotency is a characteristic that allows operation of a variable with itself without altering its value. Due to the inherent nature of asymmetric functions, true idempotency is unachievable. However, the property can theorized for the following cases of single-input operations, even though the value of the variable is not preserved. The properties are classified according to their behaviors.

\begin{identity}[Null Idempotency]
\label{id:nullIdempotency}
    	\begin{equation}
    	   \iandoption\iandtwo{A}{A}\:=\:0
    	\end{equation}
 		\begin{equation}
			 \implyoption\imptwo{A}{A}\:=\:1
		\end{equation} 
  
	\begin{proof}
		Representing the IAND operation in terms of AND, 
		\begin{equation}
		      \iandtwo{A}{A}\:=\:\andtwo{A}{\overline{A}}\:=\:0
		\end{equation}
		which proves the identity.
	\end{proof}
\end{identity}
 
\begin{identity}[Inverse Idempotency - I]
	\begin{equation}
	   \iandoption\iandtwo{A}{\overline{A}}\:=\:A
	\end{equation}
 	\begin{equation}
	   \implyoption\imptwo{A}{\overline{A}}\:=\:\overline{A}
	\end{equation}
 
	\begin{proof}
		Representing the IAND operation in terms of AND, 
		\begin{equation}
		      \iandtwo{A}{\overline{A}}\:=\:\andtwo{A}{A}\:=\:A
		\end{equation}
		substantiating the identity.
	\end{proof}
\end{identity}

\begin{identity}[Inverse-Idempotency - II]
	\begin{equation}
            \iandoption\iandtwo{\overline{A}}{A}\:=\:\overline{A}
	\end{equation}
 	\begin{equation}
            \implyoption\imptwo{\overline{A}}{A}\:=\:A
	\end{equation}
 
	\begin{proof}
		Representing the IAND operation in terms of AND, 
		\begin{equation}
		\iandtwo{\overline{A}}{A}\:=\:\andtwo{\overline{A}}{\overline{A}}\:=\:\overline{A}
		\end{equation}
		proving the identity.
	\end{proof}
\end{identity}

\subsection{Boolean Algebraic Laws}
\label{sec:booleanLaws}
The section elucidates the conventional principles of Boolean Algebra as they apply in the context of asymmetric logic functions.  It is noteworthy to mention that certain laws in this exposition may diverge from their conventional definitions; nevertheless, a discernible analogy is apparent for each.  

\subsubsection{Commutative Law}
\label{sec:commLaw}
The characterization of a Boolean function as `asymmetric' inherently denotes its departure from the conventional commutative law, \textit{i.e.}	

	\begin{equation}
	\iandtwo{A}{B}\:\neq\:\iandtwo{B}{A}
	\end{equation}
        \begin{equation}
		\hspace{-2em}\text{and, }\imptwo{A}{B}\:\neq\:\imptwo{B}{A}.
	\end{equation}

 \noindent However, a kind of asymmetric commutation is achievable as demonstrated by the following theorem. 
 
 \begin{theorem}[Asymmetric Commutation]
 \label{th:asymmcomm}
	Considering two operands $A$ and $B$, commutation is realized through complements of each of the literals as illustrated by (\ref{eq:comm1}) and (\ref{eq:icomm1})
	\begin{equation}
	\iandtwo{A}{B}\:=\:\iandtwo{\overline{B}}{\overline{A}}
	\label{eq:comm1}
	\end{equation}
         \begin{equation}
	\hspace{-2em}\text{and, }\imptwo{A}{B}\:=\:\imptwo{\overline{B}}{\overline{A}}
	\label{eq:icomm1}
	\end{equation}
 
	\begin{proof}
		Replacing the IAND with AND, 
		\begin{equation}
		\iandtwo{A}{B}\:=\:\andtwo{A}{\overline{B}}\:=\:\andtwo{\overline{B}}{A}.
		\label{eq:comm2}
		\end{equation}
		Changing the RHS of (\ref{eq:comm2}) back to IAND notation,
		\begin{equation}
		\iandtwo{A}{B}\:=\:\iandtwo{\overline{B}}{\overline{A}},
		\end{equation}
		which proves the theorem.
	\end{proof}
	\label{th:comm}
\end{theorem}

\subsubsection{Associativity Laws}
\label{sec:associativitylaw} 
Associativity laws delineate the manner in which three or more Boolean functions are paired while being operated by a Boolean operator. It is crucial to note that, owing to the asymmetric inversion of operands, IAND and IMPLY operations do not adhere to associativity in the conventional sense. 

\begin{theorem}[Conventional `Non-Associativity'] 
\label{th:nonasso}	
	$\vspace{-0.5em}$
 \begin{equation}
 				 \iandtwo{\piandtwo{A}{B}}{C}\:\neq\:\iandtwo{A}{\piandtwo{B}{C}}
 \end{equation}
		
   \noindent The above can be proved as follows. Note that the parentheses are solved two-at-a-time from left to right as suggested earlier.
	
		\begin{proof}
			Converting the two sides of Theorem \ref{th:nonasso} to AND notation,
		\begin{equation}
			\iandtwo{\piandtwo{A}{B}}{C}=\andthree{A}{\overline{B}}{\overline{C}}
			\label{eq:ass1}
		\end{equation}
			\begin{equation}
			\iandtwo{A}{\piandtwo{B}{C}}=\iandtwo{A}{\pandtwo{B}{\overline{C}}}=\andtwo{A}{(\overline{B}\vee C)}
				\label{eq:ass2}
		\end{equation}
		Clearly, (\ref{eq:ass1}) and (\ref{eq:ass2}) are not equivalent.
	\end{proof}
	
\end{theorem}

However, asymmetric functions do conform to two special kinds of associativity laws: Inverting and Non-Inverting. These are explained in detail in the following two theorems. 

\begin{theorem}[Non-Inverting Associativity]
	\label{th:asso1}
	Non-inverting associativity denotes a scenario in which altering the order of specific operands does not necessitate inversion. This circumstance arises when the positions of any two `inverted' operands are interchanged. 
	\begin{equation}
	\iand{A}{B}{C}\:=\:\iand{A}{C}{B}
	\end{equation}
	\begin{proof}
		Converting the expression to AND notation: 
		\begin{equation}
		\iand{A}{B}{C}\:=\:\andthree{A}{\overline{B}}{\overline{C}}\:=\:\andthree{A}{\overline{C}}{\overline{B}}.
		\end{equation}
		This can be rewritten using IAND notation as follows:
		\begin{equation}
		\andthree{A}{\overline{C}}{\overline{B}} \:=\: \iand{A}{C}{B}.
		\label{asso13}
		\end{equation}
		The theorem is thus proven.
	\end{proof}
\end{theorem}

\begin{theorem}[Inverting Associativity]
	\label{th:primsecond}
	Inverting associativity necessitates the inversion of both the operands that have undergone a change in position. This theorem must be employed when a `non-inverted' input interchanges its place with an `inverted' input. For two literals, the law is directly equivalent to Theorem \ref{th:asymmcomm}, where the inverted operand \textit{B} trades its position with operand \textit{A} relative to the IAND/IMPLY operator. For three literals, however, 
	\begin{equation}
	\iand{A}{B}{C}\:=\:\iand{\overline{B}}{\overline{A}}{C}\:=\:\iand{\overline{C}}{B}{\overline{A}}\:=\:\iand{\overline{C}}{\overline{A}}{B}
	\label{primsecond}
	\end{equation}
        \begin{equation}
	\imp{A}{B}{C}\:=\:\imp{\overline{C}}{B}{\overline{A}}\:=\:\imp{A}{\overline{C}}{\overline{B}} \notag \\
        \end{equation}
        \begin{equation}
        \:=\:\imp{B}{\overline{C}}{\overline{A}}
	  \label{iprimsecond}
	\end{equation}

    Note that the resultant expression on the RHS in (\ref{primsecond})  and (\ref{iprimsecond}) continue to follow the order of operation for asymmetric logic as described in section \ref{sec:asymm_func}.

	\begin{proof}
		Converting the LHS of (\ref{primsecond}) to AND notation:
		\begin{equation}
		\iand{A}{B}{C}\:=\:\andthree{A}{\overline{B}}{\overline{C}}.
		\end{equation} 
		Rearranging the inputs on the RHS of the above expression,
		\begin{equation}
		\iand{A}{B}{C}\:=\:\andthree{\overline{B}}{A}{\overline{C}}\:=\:\andthree{\overline{C}}{\overline{B}}{A}\:=\:\andthree{\overline{C}}{A}{\overline{B}},
		\label{primsecond1}
		\end{equation}
		which can be rewritten using IAND notation as 
		\begin{equation}
		\iand{A}{B}{C}\:=\:\iand{\overline{B}}{\overline{A}}{C}\:=\:\iand{\overline{C}}{B}{\overline{A}}\:=\:\iand{\overline{C}}{\overline{A}}{B},
		\label{primsecond3}
		\end{equation}
		which is the same as (\ref{primsecond}).	
	\end{proof}
\end{theorem}

\textit{Discussion:} Theorem \ref{th:asso1} and Theorem \ref{th:primsecond} can also be interpreted intuitively as the movement of inverted and non-inverted operands around the IAND/IMPLY operators:

\begin{itemize}
	\item Logical equivalence continues to be maintained if an `inverted' operand trade places with another `inverted' operand (non-inverting associativity). 
 	\item Should a `non-inverted' operand interchange positions with an `inverted' operand within the expression, both of these operands are complemented to preserve logical equivalence (inverting associativity).
\end{itemize}

\subsubsection{Distributive Laws}
\label{sec:distrlaws}
This section details the distributive laws that govern the interaction of asymmetric operators with the conventional OR and AND logic. Theorems presented here are numbered rather than named individually, since individual nomenclature would be quite challenging and perhaps, confusing. Additionally, it is noteworthy that while some of these theorems may possess greater utility or relevance than others, the entirety of the conceivable combinations of these operations are presented for completeness. Further, the proofs for each of these theorems is akin to the theorems and identities presented thus far, and is hence skipped for brevity.    

\begin{theorem}[Distributive Law - I]
	\label{th:dist1}
	\begin{equation}
	\iandtwo{A}{(B\:\wedge\:C)}\:=\:\piandtwo{A}{B}\:\vee\:\piandtwo{A}{C}.
	\label{eq:dist11}
	\end{equation}
  	\begin{equation}
	\imptwo{A}{(B\:\wedge\:C)}\:=\:\pimptwo{A}{B}\:\wedge\:\pimptwo{A}{C}.
	\label{idist}
	\end{equation}
 \end{theorem}

\begin{theorem}[Distributive Law - II]
	\label{th:dist2}
	\begin{equation}
		\andtwo{(\iandtwo{A}{B})}{C}\:=\:\iand{A}{B}{\overline{C}}
		\label{eq:dist21}
	\end{equation}
         \begin{equation}
	\andtwo{(\imptwo{A}{B})}{C}\:=\:\pandtwo{\overline{A}}{C} \vee \pandtwo{B}{C}
	\label{eq:idist21}
	\end{equation}
\end{theorem}

\begin{theorem}[Distributive Law - III]
\label{th:dist3}
	\begin{equation}
	\iandtwo{(A \vee B)}{C}\:=\:\piandtwo{A}{C}\:\vee\:\piandtwo{B}{C}
	\label{eq:dist31}
	\end{equation}
         \begin{equation}
	\imptwo{(A \vee B)}{C}\:=\:\pimptwo{A}{C}\:\wedge\:\pimptwo{B}{C}
	\label{iordist}
	\end{equation}
\end{theorem}

\begin{theorem}[Distributive Law - IV] 
	\begin{equation}
	\iandtwo{A}{(B \vee C)}\:=\: \andtwo{(\iandtwo{A}{B})}{(\iandtwo{A}{C})}
	\label{eq:dist51}
	\end{equation}
        \begin{equation}
	\imptwo{A}{(B \vee C)} = \pimptwo{A}{B} \vee C = A \rightarrow (\overline{B} \rightarrow C)
	\label{eq:idist51}
	\end{equation}
\end{theorem}

\begin{theorem}[Distributive Law - V]
	\label{th:dist7}
	\begin{equation}
	\andtwo{A}{(\iandtwo{B}{C})}\:=\:\iandtwo{(A \wedge B)}{C}\:=\:\iand{A}{\overline{B}}{C}
	\label{eq:dist71}
	\end{equation}
         \begin{equation}
	\andtwo{A}{(\imptwo{B}{C})} = \pandtwo{A}{\overline{B}} \vee \pandtwo{A}{C}
	\label{eq:idist61}
	\end{equation}

 \end{theorem}

\begin{theorem}[Distributive Law - VI]
	\begin{equation}
	A \vee \piandtwo{B}{C}\:=\:\andtwo{(A \vee B)}{(A \vee \overline{C})}
	\label{eq:dist41}
	\end{equation} 
  \begin{equation}
	A \vee \pimptwo{B}{C}\:=\:\overline{A} \rightarrow (B \rightarrow C)
	\label{eq:idist41}
	\end{equation} 
 \end{theorem}

\begin{theorem}[Distributive Law - VII] 
	\begin{equation}
	\piandtwo{A}{B} \vee C\:=\: \andtwo{(A \vee C)}{(\overline{B} \vee C)}
	\label{eq:dist61}
	\end{equation}
        \begin{equation}
	\imptwo{(A \wedge B)}{C}\:=\: A \rightarrow (B \rightarrow C)
	\label{eq:idist71}
	\end{equation}
 \end{theorem}


\subsubsection{De Morgan's Law for Asymmetric Logic}
\label{sec:demorgan}
By now it is evident that traditional Boolean laws do not apply to asymmetric logic operations without requisite modifications. Fortunately, for the De Morgan's rule, there is a notable resemblance between the proposed modified version for IAND/IMPLY logic and the established convention. The De Morgan's law as applied to asymmetric logic operations can be formally stated as,

\begin{theorem}
\label{th:dm}
(A) The negation of ordered IAND of two literals is equal to the OR of the two literals with the non-inverted term complemented. \\ \\
(B) The negation of ordered implication of two literals is equal to the NAND of the two literals with the non-inverted term complemented. 

\flushleft For two inputs:
	\begin{equation}
		\iandoption\overline{\piandtwo{A}{B}} = \overline{A} \vee B
		\label{eq:dm1}
	\end{equation}
         \begin{equation}
         \label{eq:dm2}
         \implyoption\overline{\pimptwo{A}{B}} = A \wedge \overline{B}
         \end{equation}
         Conversely,
	 \begin{equation}
		 \iandoption\overline{(A \vee B)} = \iandtwo{\overline{A}}{B}
		\label{eq:dm3}
	\end{equation}
 	\begin{equation}
	\implyoption\overline{\pandtwo{A}{B}} = \imptwo{A}{\overline{B}}
	\label{eq:idm12}
	\end{equation}
	For three inputs:
	\begin{equation}
		\iandoption\overline{\piand{A}{B}{C}} = \overline{A} \vee B \vee C
		\label{eq:dm4}	
	\end{equation}
 	\begin{equation}
	\implyoption\overline{\pimp{A}{B}{C}} = A \wedge B \wedge \overline{C}
	\label{eq:idm2}	
	\end{equation}
 Conversely, 
	\begin{equation}
		\iandoption\overline{(A \vee B \vee C)} = \iand{\overline{A}}{B}{C}
		\label{eq:dm5}	
	\end{equation}
 	\begin{equation}
	\implyoption\overline{\pandthree{A}{B}{C}} = \imp{A}{B}{\overline{C}}
	\label{eq:idm22}	
	\end{equation}

For memristive circuits, the De-Morgan's rule can be restated in terms of the fundamental IMPLY/NAND logic set as follows. Note that the inverter can easily be realized using a NAND gate with inputs shorted.

	\begin{equation}
	\overline{\pimptwo{A}{B}} = \overline{\overline{A \wedge \overline{B}}} = \mathrm{INV \;( NAND\;(A, \overline{B}))}
	\label{eq:idm3}
	\end{equation}
	
and,
	\begin{equation}
	\overline{\pimp{A}{B}{C}} = \overline{ \overline{A \wedge B \wedge \overline{C}} } = \mathrm{INV \;( NAND\;(A,B, \overline{C}))}
	\label{eq:idm4}	
	\end{equation}

Although the law is stated for two and three inputs only, its applicability can be expanded to accommodate any number of inputs. In each instance, the non-inverted input will consistently appear as its own complement accompanied by the modification in operators as described above. 
\end{theorem}

\subsubsection{Principle of Duality}

Traditionally, the `dual' of a Boolean function is derived by replacing the ANDs (ORs) and 1s (0s) with ORs (ANDs) and 0s (1s), respectively. For instance, the dual of $(A \vee B)$ is $(A \wedge B)$, and $(D \vee 1)$ is the dual of $(D \wedge 0)$. 

\begin{theorem}
Boolean duals for expressions involving asymmetric logic can be found by following the below steps in order:
\begin{itemize}
	\item Replace all ANDs with ORs, and vice versa.
	\item Replace all 1s with 0s, and vice versa.
	\item Change all IANDs to ORs with all inverted inputs complemented.
	\item Replace all IMPLY with ANDs and complement all of the inverted inputs.  
\end{itemize}  
\end{theorem}
\noindent For example, the dual of the Boolean expression $\iand{A}{B}{C}$ is $(A \vee \overline{B} \vee \overline{C})$, and that of $\imp{A}{B}{C}$ becomes $(\overline{A} \vee \overline{B} \vee C)$.
 

\subsection{Canonical Normal Forms for Asymmetric Logic}

Boolean Algebra defines two types of canonical normal forms: the canonical disjunctive normal form (CDNF) and the canonical conjunctive normal form (CCNF). The IAND/OR and IMPLY/NAND are fundamental logic sets capable of representing any Boolean function. Hence, it is vital to define the canonical normal forms for these logic sets. 

\subsubsection{Canonical Disjunctive Normal Form (CDNF)} 
CDNF, commonly referred to as the minterm canonical form or canonical sum of products (SOP), is a Boolean expression derived through the logical `OR-ing' (disjunction) of specific `AND-ed' (conjunction) Boolean literals. (\ref{eq:cdnf1}) presents an example. 
\begin{equation}
		f_{sop}(A,B,C)=\pandthree{A}{B}{C} \vee \pandthree{A}{\overline{B}}{\overline{C}}
			\label{eq:cdnf1}
\end{equation}
In the context of the IAND/OR logic set, the CDNF is proposed as the Sum of IANDs (SOI), characterized by the logical OR operation applied to specific Boolean literals that are combined through the IAND operation in various configurations. For instance,
\begin{equation}
   f_{soi}(A,B,C)=\piand{A}{B}{C} \vee \piand{A}{\overline{B}}{\overline{C}}. 
	\label{eq:cdnf2}
\end{equation} 
For the IMPLY/NAND logic set, the CDNF is proposed as canonical NAND of implication (NOI), which is the NAND of Boolean literals IMPLY-ed in different combinations. For example,
\begin{equation}
\label{eq:icdnf2}
f_{noi}(A, B, C)=\overline{\pimp{\overline{A}}{B}{C}\wedge\pimp{A}{B}{C}}.
\end{equation} 

\subsubsection{Canonical conjunctive normal form (CCNF)} 
\label{sec:ccnf}
CCNF is essentially the dual of CDNF and is also known as a maxterm or product of sums (POS). For example,
\begin{equation}
f_{pos}(A,B,C)=\andtwo{(A \vee B \vee C)}{(A \vee \overline{B} \vee C)}
\end{equation}
CCNF for the IAND/OR and IMPLY/NAND logic sets is proposed as IAND of sums (IOS) and IMPLY of NANDs (ION) respectively. Examples for each of these are presented below:
\begin{equation}
f_{ios}(A,B,C)=\iandtwo{(A \vee B \vee C)}{(\overline{A} \vee B \vee C)}
\label{eq:ccnf2}
\end{equation} 
\begin{equation}
f_{ion}(A,B,C)=\imptwo{(\overline{\andthree{A}{B}{C}})}{(\overline{\andthree{\overline{A}}{B}{C}})}
\label{eq:iccnf2}
\end{equation} 

\subsection{Relationship between IAND and Implication Logic (De Morgan Duality)}
\label{sec:relationship}

Further analysis of the De Morgan's law for asymmetric logic function reveals an interesting relationship between IAND and IMPLY operations. Per (\ref{eq:dm2}):

	\begin{equation}
	\overline{\pimptwo{A}{B}} = \andtwo{A}{\overline{B}}
	\label{eq:rel1}
	\end{equation} 
 The RHS of the above equation can be expressed as an IAND of $A$ and $B$, 
	\begin{equation}
	\overline{\pimptwo{A}{B}} = \iandtwo{A}{B}
	\label{eq:rel2}
	\end{equation}
Refactoring LHS of (\ref{eq:rel2}) as per Theorem \ref{th:asymmcomm},
	\begin{equation}
	\overline{\pimptwo{A}{B}} = \iandtwo{\overline{B}}{\overline{A}}
	\label{eq:rel2_1}
	\end{equation}
For three literals, it can be demonstrated that
	\begin{equation}
	\overline{\pimp{A}{B}{C}}  = \iand{\overline{C}}{\overline{B}}{\overline{A}}.
	\label{eq:rel4}
	\end{equation}
The converse of (\ref{eq:rel1}) and (\ref{eq:rel4}) are also true,
	\begin{equation}
	\overline{\piandtwo{A}{B}} = \imptwo{\overline{B}}{\overline{A}},
	\label{eq:rel5}
	\end{equation}
 and,
	\begin{equation}
	\overline{\piand{A}{B}{C}}  = \imp{\overline{C}}{\overline{B}}{\overline{A}}.
	\label{eq:rel6}
	\end{equation}

The De Morgan dual for any conventional logic operation can be calculated as
\begin{equation}
	f_d(a_1,a_2,a_3...a_n) = \overline{f(\overline{a_1},\overline{a_2},\overline{a_3}...\overline{a_n})}.
	\label{eq:rel7}
\end{equation} 
Remarkably, observations from (\ref{eq:rel1})-(\ref{eq:rel6}) reveal the general expression for evaluating the De Morgan dual for any IAND/IMPLY Boolean function: 
\begin{equation}
f_d(a_1,a_2,a_3...a_n) = \overline{f(\overline{a_n},\overline{a_{n-1}},\overline{a_{n-2}}...\overline{a_2},\overline{a_1})}
\label{eq:rel8}
\end{equation}
where $f_d$ and $f$ are IAND and IMPLY functions respectively (or vice versa). As noted earlier in the paper, the order of operation must be maintained when applying the above relationship. Considering the concepts and observations presented in this section along with section (\ref{sec:demorgan}), it can be asserted that IAND and IMPLY operations are "De Morgan duals" of each other. This relationship also facilitates the conversion of an SOI expression to its equivalent NOI (or vice versa). 

\section{Conclusion}
\label{sec:conclusion}
In this paper, we presented a comprehensive set of algebraic identities and theorems customized for asymmetric logic functions, particularly focusing on the logic sets fundamental to the bilayer avalanche spin-diode and memristor. This algebra marks a significant step towards optimizing computation for emerging technologies reliant on such functions by providing a foundational framework for optimizing logic circuits without conversion to and from conventional Boolean algebra. By leveraging the unique properties of non-commutative IAND and IMPLY operations, we have shown significant computational advantages, exemplified by the 28\% reduction in computational step count for a memristive full adder circuit. It is clear that embracing the inherent asymmetry of certain logic functions and developing specialized algebraic tools that harness their potential can pave the way for more efficient and advanced logic design paradigms in the era of post-CMOS computing.


\bibliographystyle{myIEEEtran_X}
\bibliography{Equations.bib}

\begin{thebibliography}{10}
\providecommand{\url}[1]{#1}
\csname url@samestyle\endcsname
\providecommand{\newblock}{\relax}
\providecommand{\bibinfo}[2]{#2}
\providecommand{\BIBentrySTDinterwordspacing}{\spaceskip=0pt\relax}
\providecommand{\BIBentryALTinterwordstretchfactor}{4}
\providecommand{\BIBentryALTinterwordspacing}{\spaceskip=\fontdimen2\font plus
\BIBentryALTinterwordstretchfactor\fontdimen3\font minus
  \fontdimen4\font\relax}
\providecommand{\BIBforeignlanguage}[2]{{%
\expandafter\ifx\csname l@#1\endcsname\relax
\typeout{** WARNING: IEEEtran.bst: No hyphenation pattern has been}%
\typeout{** loaded for the language `#1'. Using the pattern for}%
\typeout{** the default language instead.}%
\else
\language=\csname l@#1\endcsname
\fi
#2}}
\providecommand{\BIBdecl}{\relax}
\BIBdecl

\bibitem{kim2003leakage}
N.~S. Kim, T.~Austin, D.~Blaauw, T.~Mudge, K.~Flautner, J.~S. Hu, M.~J. Irwin,
  M.~Kandemir, and V.~Narayanan, ``Leakage current: Moore's law meets static
  power,'' \emph{computer}, vol.~36, no.~12, pp. 68--75, 2003.

\bibitem{lundstrom2022moore}
M.~S. Lundstrom and M.~A. Alam, ``Moore’s law: The journey ahead,''
  \emph{Science}, vol. 378, no. 6621, pp. 722--723, 2022.

\bibitem{bohr2017cmos}
M.~T. Bohr and I.~A. Young, ``Cmos scaling trends and beyond,'' \emph{IEEE
  Micro}, vol.~37, no.~6, pp. 20--29, 2017.

\bibitem{waldrop2016chips}
M.~M. Waldrop, ``The chips are down for moore’s law,'' \emph{Nature News},
  vol. 530, no. 7589, p. 144, 2016.

\bibitem{finocchio2023roadmap}
G.~Finocchio, J.~A.~C. Incorvia, J.~S. Friedman, Q.~Yang, A.~Giordano,
  J.~Grollier, H.~Yang, F.~Ciubotaru, A.~Chumak, A.~Naeemi \emph{et~al.},
  ``Roadmap for unconventional computing with nanotechnology,'' \emph{Nano
  Futures}, 2023.

\bibitem{liu2019hybrid}
W.~Liu, P.~K.~J. Wong, and Y.~Xu, ``Hybrid spintronic materials: Growth,
  structure and properties,'' \emph{Progress in Materials Science}, vol.~99,
  pp. 27--105, 2019.

\bibitem{lin2019two}
X.~Lin, W.~Yang, K.~L. Wang, and W.~Zhao, ``Two-dimensional spintronics for
  low-power electronics,'' \emph{Nature Electronics}, vol.~2, no.~7, pp.
  274--283, 2019.

\bibitem{bromberg2012novel}
D.~M. Bromberg, D.~H. Morris, L.~Pileggi, and J.-G. Zhu, ``Novel stt-mtj device
  enabling all-metallic logic circuits,'' \emph{IEEE transactions on
  Magnetics}, vol.~48, no.~11, pp. 3215--3218, 2012.

\bibitem{cmatTED}
J.~S. Friedman and A.~V. Sahakian, ``{Complementary magnetic tunnel junction
  logic},'' \emph{IEEE Transactions on Electron Devices}, vol.~61, no.~4, pp.
  1207--1210, 2014.

\bibitem{allwood2005magnetic}
D.~A. Allwood, G.~Xiong, C.~Faulkner, D.~Atkinson, D.~Petit, and R.~Cowburn,
  ``Magnetic domain-wall logic,'' \emph{Science}, vol. 309, no. 5741, pp.
  1688--1692, 2005.

\bibitem{omari2014chirality}
K.~Omari and T.~Hayward, ``Chirality-based vortex domain-wall logic gates,''
  \emph{Physical Review Applied}, vol.~2, no.~4, p. 044001, 2014.

\bibitem{dml3}
P.~Xu, K.~Xia, C.~Gu, L.~Tang, H.~Yang, and J.~Li, ``An all-metallic logic gate
  based on current-driven domain wall motion.'' \emph{Nature nanotechnology},
  vol.~3, no.~2, 2008.

\bibitem{dml4_mqca}
\BIBentryALTinterwordspacing
A.~Imre, G.~Csaba, L.~Ji, A.~Orlov, G.~H. Bernstein, and W.~Porod, ``Majority
  logic gate for magnetic quantum-dot cellular automata,'' \emph{Science}, vol.
  311, no. 5758, pp. 205--208, 2006.
\BIBentrySTDinterwordspacing

\bibitem{allcarbon}
J.~S. Friedman, A.~Girdhar, R.~M. Gelfand, G.~Memik, H.~Mohseni, A.~Taflove,
  B.~W. Wessels, J.-P. Leburton, and A.~V. Sahakian, ``Cascaded spintronic
  logic with low-dimensional carbon,'' \emph{Nature communications}, vol.~8, p.
  15635, 2017.

\bibitem{walker2023near}
B.~W. Walker, A.~J. Edwards, X.~Hu, M.~P. Frank, F.~Garcia-Sanchez, and J.~S.
  Friedman, ``Near-landauer reversible skyrmion logic with voltage-based
  propagation,'' \emph{arXiv preprint arXiv:2301.10700}, 2023.

\bibitem{datta1990electronic}
S.~Datta and B.~Das, ``Electronic analog of the electro-optic modulator,''
  \emph{Applied Physics Letters}, vol.~56, no.~7, pp. 665--667, 1990.

\bibitem{spinfet2_pre}
\BIBentryALTinterwordspacing
H.~C. Koo, J.~H. Kwon, J.~Eom, J.~Chang, S.~H. Han, and M.~Johnson, ``Control
  of spin precession in a spin-injected field effect transistor,''
  \emph{Science}, vol. 325, no. 5947, pp. 1515--1518, 2009.
\BIBentrySTDinterwordspacing

\bibitem{spinfet3_nb}
J.~Schliemann, J.~C. Egues, and D.~Loss, ``Nonballistic spin-field-effect
  transistor,'' \emph{Physical review letters}, vol.~90, no.~14, p. 146801,
  2003.

\bibitem{spinfet4_q}
B.~Wang, J.~Wang, and H.~Guo, ``Quantum spin field effect transistor,''
  \emph{Physical Review B}, vol.~67, no.~9, p. 092408, 2003.

\bibitem{barla2021spintronic}
P.~Barla, V.~K. Joshi, and S.~Bhat, ``Spintronic devices: a promising
  alternative to cmos devices,'' \emph{Journal of Computational Electronics},
  vol.~20, no.~2, pp. 805--837, 2021.

\bibitem{basdljf}
\BIBentryALTinterwordspacing
J.~S. Friedman, E.~R. Fadel, B.~W. Wessels, D.~Querlioz, and A.~V. Sahakian,
  ``Bilayer avalanche spin-diode logic,'' \emph{AIP Advances}, vol.~5, no.~11,
  p. 117102, 2015.
\BIBentrySTDinterwordspacing

\bibitem{vyas2018sequential}
V.~Vyas and J.~S. Friedman, ``Sequential circuit design with bilayer avalanche
  spin diode logic,'' in \emph{Proceedings of the 14th IEEE/ACM International
  Symposium on Nanoscale Architectures}, 2018, pp. 49--50.

\bibitem{kvatinsky2011memristor}
S.~Kvatinsky, A.~Kolodny, U.~C. Weiser, and E.~G. Friedman, ``Memristor-based
  imply logic design procedure,'' in \emph{2011 IEEE 29th International
  Conference on Computer Design (ICCD)}.\hskip 1em plus 0.5em minus 0.4em\relax
  IEEE, 2011, pp. 142--147.

\bibitem{Kvatinsky2014}
S.~Kvatinsky, G.~Satat, N.~Wald, E.~G. Friedman, and U.~C. Kolodny,
  A.and~Weiser, ``{Memristor-based material implication (IMPLY) logic: Design
  principles and methodologies},'' \emph{IEEE Transactions on Very Large Scale
  Integration (VLSI) Systems}, vol.~22, no.~10, pp. 2054--2066, 2014.

\bibitem{Chattopadhyay2011}
A.~Chattopadhyay and Z.~Rakosi, ``{Combinational logic synthesis for material
  implication},'' \emph{2011 IEEE/IFIP 19th International Conference on VLSI
  and System-on-Chip, VLSI-SoC 2011}, pp. 200--203, 2011.

\bibitem{Raghuvanshi2014}
\BIBentryALTinterwordspacing
A.~Raghuvanshi and M.~Perkowski, ``Logic synthesis and a generalized notation
  for memristor-realized material implication gates,'' in \emph{Proceedings of
  the 2014 IEEE/ACM International Conference on Computer-Aided Design}, ser.
  ICCAD '14.\hskip 1em plus 0.5em minus 0.4em\relax Piscataway, NJ, USA: IEEE
  Press, 2014, pp. 470--477.
\BIBentrySTDinterwordspacing

\bibitem{Lehtonen2012}
E.~Lehtonen, J.~Poikonen, and M.~Laiho, ``Implication logic synthesis methods
  for memristors,'' in \emph{2012 IEEE International Symposium on Circuits and
  Systems}, May 2012, pp. 2441--2444.

\bibitem{Shirinzadeh2016}
S.~Shirinzadeh, M.~Soeken, and R.~Drechsler, ``{Multi-objective BDD
  optimization for RRAM based circuit design},'' in \emph{Formal Proceedings of
  the 2016 IEEE 19th International Symposium on Design and Diagnostics of
  Electronic Circuits and Systems, DDECS 2016}, 2016.

\bibitem{Marranghello2015a}
\BIBentryALTinterwordspacing
F.~S. Marranghello, V.~Callegaro, A.~I. Reis, and R.~P. Ribas, ``{SOP based
  logic synthesis for memristive IMPLY stateful logic},'' in \emph{2015 33rd
  IEEE International Conference on Computer Design (ICCD)}.\hskip 1em plus
  0.5em minus 0.4em\relax IEEE, oct 2015, pp. 228--235.
\BIBentrySTDinterwordspacing

\bibitem{Chakraborti2014}
S.~Chakraborti, P.~V. Chowdhary, K.~Datta, and I.~Sengupta, ``Bdd based
  synthesis of boolean functions using memristors,'' in \emph{Design \& Test
  Symposium (IDT), 2014 9th International}.\hskip 1em plus 0.5em minus
  0.4em\relax IEEE, 2014, pp. 136--141.

\bibitem{Lalchhandama2016}
\BIBentryALTinterwordspacing
F.~Lalchhandama, B.~G. Sapui, and K.~Datta, ``{An improved approach for the
  synthesis of Boolean functions using memristor based IMPLY and INVERSE-IMPLY
  gates},'' in \emph{2016 IEEE Computer Society Annual Symposium on VLSI
  (ISVLSI)}, vol. 2016-Septe.\hskip 1em plus 0.5em minus 0.4em\relax IEEE, jul
  2016, pp. 319--324.
\BIBentrySTDinterwordspacing

\bibitem{Fan2014}
D.~Fan and K.~Sharad, M.and~Roy, ``{Design and synthesis of ultralow energy
  spin-memristor threshold logic},'' \emph{IEEE Transactions on
  Nanotechnology}, vol.~13, no.~3, pp. 574--583, 2014.

\bibitem{vyaskmap}
V.~Vyas, L.~Jiang-Wei, P.~Zhou, X.~Hu, and J.~S. Friedman, ``Karnaugh map
  method for memristive and spintronic asymmetric basis logic functions,''
  \emph{IEEE Transactions on Computers}, vol.~70, no.~1, pp. 128--138, 2020.

\bibitem{memristorchua}
L.~Chua, ``Memristor-the missing circuit element,'' \emph{IEEE Transactions on
  Circuit Theory}, vol.~18, no.~5, pp. 507--519, September 1971.

\bibitem{Strukov2009}
\BIBentryALTinterwordspacing
D.~B. Strukov, G.~S. Snider, D.~R. Stewart, and R.~S. Williams, ``{The missing
  memristor found},'' \emph{Nature}, vol. 459, no. 7250, pp. 1154--1154, 2009.
\BIBentrySTDinterwordspacing

\bibitem{Borghetti2010}
\BIBentryALTinterwordspacing
J.~Borghetti, G.~S. Snider, P.~J. Kuekes, J.~J. Yang, D.~R. Stewart, and R.~S.
  Williams, ``{'Memristive' switches enable 'stateful' logic operations via
  material implication},'' \emph{Nature}, vol. 464, no. 7290, pp. 873--876,
  2010.
\BIBentrySTDinterwordspacing

\end{thebibliography}

\begin{IEEEbiography}
[{\includegraphics[width=0.95in,clip,keepaspectratio]{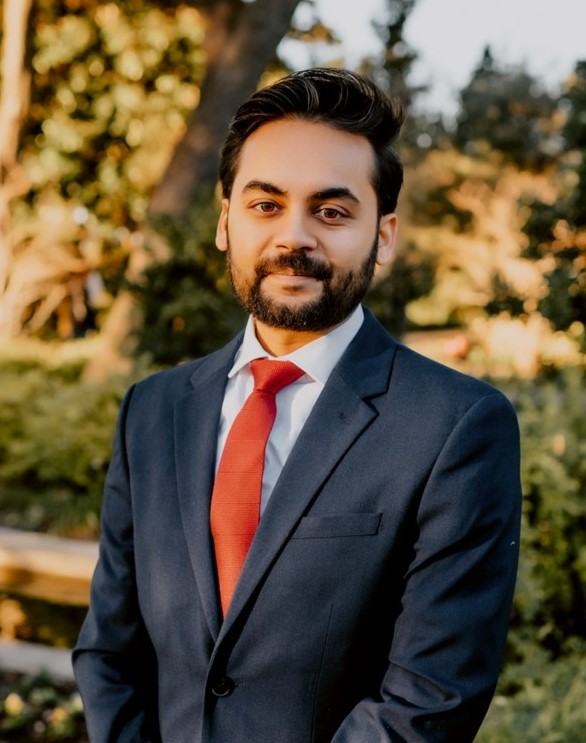}}]
{Vaibhav Vyas} completed his M.S. in Electrical Engineering in 2018 from the Erik Jonsson School of Engineering \& Computer Science at The University of Texas at Dallas. Prior to this, he earned his B.E. degree in Electronics and Communication Engineering in 2016 from Rajiv Gandhi Proudyogiki Vishwavidyalaya (R.G.P.V.) University. His research interests lie in the domain of logic design automation for emerging device technologies. Following his graduation from UT Dallas, Vaibhav gained professional experience as an Embedded Software Engineer in the Precision Ag Department at John Deere Intelligent Solutions Group, Des Moines (IA). In September 2019, he transitioned to the Firmware Engineering Team at Extron Electronics, where he contributes to the development of a diverse range of Extron products.
\end{IEEEbiography}
\vspace{-27em}
\begin{IEEEbiography}
[{\includegraphics[width=1in,height=1.25in,clip,keepaspectratio]{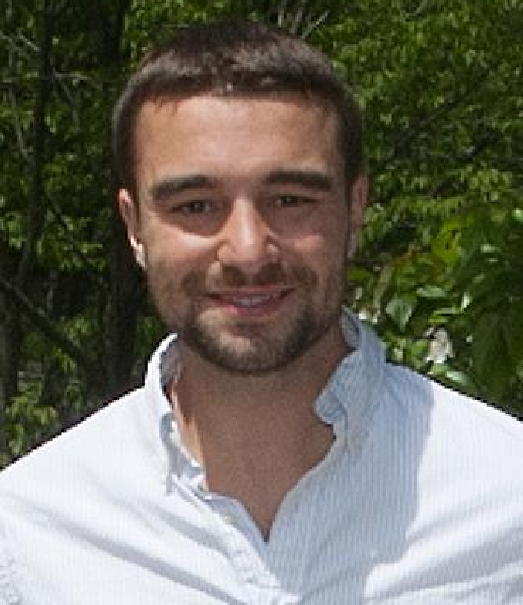}}]
{Joseph S. Friedman} (S'09–M'14-SM'19) received the A.B. and B.E. degrees from Dartmouth College, Hanover, NH, USA, in 2009, and the M.S. and Ph.D. degrees in electrical \& computer engineering from Northwestern University, Evanston, IL, USA, in 2010 and 2014, respectively. \par
He joined The University of Texas at Dallas, Richardson, TX, USA, in 2016, where he is currently an Associate Professor of electrical \& computer engineering, a core member of the Computer Engineering program, and director of the NeuroSpinCompute Laboratory. From 2014 to 2016, he was a Centre National de la Recherche Scientifique Research Associate with the Institut d'Electronique Fondamentale, Universit\'{e} Paris-Sud, Orsay, France. He has also been a Summer Faculty Fellow at the U.S. Air Force Research Laboratory, Rome, NY, USA, a Visiting Professor at Politecnico di Torino, Turin, Italy, a Guest Scientist at RWTH Aachen University, Aachen, Germany, and worked on logic design automation as an intern at Intel Corporation, Santa Clara, CA, USA.\par
Dr. Friedman is a member of the editorial boards of \textit{Scientific Reports} and \textit{IEEE Transactions on Nanotechnology}, and previously the \textit{Microelectronics Journal}. He is the general chair of the IEEE International Conference on Rebooting Computing (ICRC), conference chair of SPIE Spintronics, has served on numerous conference technical program committees, and is the founder and chairperson of the Texas Symposium on Computing with Emerging Technologies (ComET). He has also been awarded the National Science Foundation (NSF) Faculty Early Career Development Program (CAREER) Award. His research interests include the invention and design of novel logical and neuromorphic computing paradigms based on nanoscale and quantum mechanical phenomena, with particular emphasis on spintronics.\par
\end{IEEEbiography}

\end{document}